# Long-distance frequency dissemination with a resolution of $10^{-17}$

C. DAUSSY, O. LOPEZ[*], A. AMY-KLEIN[*], A. GONCHAROV[♦], M. GUINET, C. CHARDONNET
LPL, Laboratoire de Physique des Lasers, UMR 7538 C.N.R.S., Université Paris 13, 99, av. J.-B. Clément, 93430 Villetaneuse, France.

F. NARBONNEAU, M. LOURS, D. CHAMBON, S. BIZE, A. CLAIRON, G. SANTARELLI
SYRTE, Systèmes de Référence Temps Espace, UMR 8630 CNRS, Observatoire de Paris, 61 avenue de l'Observatoire 75014 Paris, France.

M.E TOBAR, A.N. LUITEN
School of Physics, University of Western Australia, 35 Stirling Hwy, Crawley, WA

ABSTRACT

We use a new technique to disseminate microwave reference signals along ordinary optical fiber. The fractional frequency resolution of a link of 86 km in length is $10^{-17}$ for a one day integration time, a resolution higher than the stability of the best microwave or optical clocks. We use the link to compare the microwave reference and a $CO_2/OsO_4$ frequency standard that stabilizes a femtosecond laser frequency comb. This demonstrates a resolution of $3\times10^{-14}$ at 1 s. An upper value of the instability introduced by the femtosecond laser-based synthesizer is estimated as $1\times10^{-14}$ at 1 s.

PACS : 06.20.-f, 42.62.Eh, 06.30.Ft

---

[*] Corresponding author : fax: +33 (0)1 49 40 32 00, e-mail: amy@galilee.univ-paris13.fr
[♦] Permanent address : Institute of Laser Physics, Novosibirsk, Russia, fax: +7 3832 33 20 67

Significant progress has been made in the past few years with the development of atomic and ion optical frequency standards. For example, clocks are presently being built based on $Hg^+$ [1] and $Yb^+$ ions [2], as well as Ca [3] or Sr [4] atoms. These projects are aiming to reach an accuracy in the range of $10^{-17}$-$10^{-18}$ in the near future. In this context the precise comparison of distant frequency standards becomes a crucial issue, since a typical optical clock is not portable. Such systems will also be extremely important for those future tests of fundamental physics that are based on accurate time and frequency comparisons. Examples of this include the search for possible time variations of fundamental constants [5] and tests of Lorentz Invariance [6].

Distant clock comparisons can be performed using a satellite link, such as GPS, giving a resolution at 1 day of a few $10^{-14}$ with a commercial receiver, or a few $10^{-15}$ using a geodesic receiver, or two-way satellite time transfer systems (TWSTT). Alternatively, clock comparisons can be made indirectly using transportable clocks as transfer oscillators between fixed high-stability clocks. However, in all but a few cases the performance of transportable clocks is significantly worse than the best available frequency standards, which are complex devices that are difficult to move from their laboratories. Presently the best of the transportable clocks show an accuracy of slightly below $10^{-15}$ for the Cs fountain [7], $2\times10^{-13}$ for a transportable $CH_4$ standard [8, 9], and $1.3\times10^{-12}$ for a Ca transportable standard [10]. For more distant laboratories optical fibers can be used to disseminate the RF signal of the primary standard, or directly distribute the optical frequency signal, as demonstrated for the 3.5 km link between NIST and JILA, with a resolution of $4\times10^{-14}$ at 1 s [11]. In addition, the advent of femtosecond (fs) laser-based optical frequency synthesizers have greatly simplified the frequency comparison of different optical frequency signals. For instance a resolution of $7\times10^{-15}$ at 1 s was recently obtained for a frequency comparison between $Hg^+$ and Ca standard [12]. It was also established that the fractional frequency uncertainty introduced by the fs laser synthesizer can be in the $10^{-17}$ domain at 1 s and is approaching the $10^{-19}$ domain after 100 s [13] when comparing two optical signals. However, the sensitivity and accuracy of optical-microwave frequency comparisons is often limited by noise processes which occur in the microwave frequency domain. For instance the upper limit of the instability of a fs laser-based synthesizer that connects the microwave and optical domains has been reported to $2\times10^{-14}$ at 1 s [14].

In this paper we demonstrate, for the first time, that two standards can be compared over a distance of 43 km, using a standard telecommunication optical link, with a resolution of $8\times10^{-15}$ at 1 s, and $10^{-17}$ at one day integration time. As an application, we measured the frequency of a stabilized $CO_2$ laser at Laboratoire de Physique des Lasers (LPL) against the primary frequency standard of the Systèmes de Référence Temps Espace (SYRTE), with a resolution of $3\times10^{-14}$ at 1 s integration time, limited by the $CO_2$ laser frequency stability. From this measurement, the instabilities of the fs laser-based synthesizer connecting the microwave and optical domains can be estimated to be less than $1\times10^{-14}$ at 1 s.

The LPL, located on the outskirts of Paris, develops optical frequency standards in the mid infrared. The SYRTE, in the centre of Paris, develops Cs atomic fountains with an accuracy of $7\times10^{-16}$, and a frequency stability of $1.6\times10^{-14}\tau^{-1/2}$ [15]. A 100MHz high stability flywheel oscillator has been developed at SYRTE, which is based on a combination of a cryogenic Sapphire oscillator (CSO), an H-Maser and a set of low noise microwave synthesizers [16]. The conventional means of characterizing the frequency stability of an oscillator is the Allan frequency deviation [17]. It is given by: $\sigma_y(\tau) = \sqrt{\frac{1}{2}\left\langle\left(\bar{y}_n - \bar{y}_{n+1}\right)^2\right\rangle}$, where $y_n$ is the nth measurement of the fractional average frequency of the oscillator measured over an integration time, $\tau$. The flywheel oscillator shows a frequency stability slightly below $10^{-14}\times\tau^{-1}$ in the range 1-10 s, and $1\text{-}2\times 10^{-15}$ from 10 to $10^5$ s (see Allan deviation as dashed line in Fig. 1). The frequency of this signal is steered by the H-Maser in the long-term, and monitored by the Cs atomic fountain.

Two parallel optical links have been used to transmit the reference signal between SYRTE and LPL [18]. A few sections of standard 1.55-µm telecommunication single mode optical fibers have been interconnected by splicing leading to a total length of 43 km. The optical link is fed with a distributed feedback semiconductor laser emitting at 1.55 µm. The junction current of the diode is modulated with the 100 MHz reference signal, which generates synchronous intensity modulation of its output light. The intensity modulation is then detected at the distant end of the optical link. The fibers are sensitive to both acoustic noise and temperature variations. A preliminary measurement of the metrological features of this link was made by redirecting the output signal from one fiber back into the second fiber leading to a complete round trip (86-km) of the 100 MHz signal [18]. The phase difference between the input and return signal was detected and analyzed in terms of frequency fluctuations. In a second experiment, the same 100 MHz signal was sent in both fibers, and the two output signals were compared. Their phase noise fluctuations were found to be of a few $10^{-15}$ at 1 s, which is 10 times less than the noise seen on a single fiber. This demonstrates that the phase fluctuations arising in both fibers are strongly correlated, and that the Allan frequency deviation for a one-way pass can be deduced by dividing by 2 the Allan deviation of the round trip, at least on time scales longer than 1 s. The Allan deviation for complete round trip in a single fiber is reported in Fig. 1 (up triangles). The typical short-term noise is $3 \times 10^{-14}$ at 1 s, while the long-term is dominated by diurnal temperature variations of about $3-5 \times 10^{-15}$. This performance, remarkable for a non-dedicated fiber, is occasionally degraded by noise bursts. For precise metrological applications, one must implement a method for compensation of the fiber phase noise.

In order to fully characterize the frequency transfer we have implemented two independent correction systems on the two fibers. Figure 2 displays a schematic of the measurement set-up. A first compensation system is used to correct the perturbation introduced on the RF reference signal through the optical link from SYRTE to LPL. The corrected signal received at LPL is sent back to SYRTE for evaluation via the second fiber. This transfer requires a second correction system. By comparing the reference with the return signal a complete stability analysis can be done.

The first compensation system implemented at SYRTE, is based on RF signal processing, and is very similar to the design used at Jet Propulsion Laboratory for the Deep Space Network distribution [19]. The RF signal detected at SYRTE after one round trip in the same fiber is compared to the reference signal. The perturbation value is extracted and used to compensate one-path perturbations by phase shifting the 100 MHz signal in the opposite sense in order to get a corrected 100 MHz signal at LPL.

The second correction system, implemented at LPL, acts directly on the optical path. A correction signal is generated by comparing the phase of the 100 MHz arriving from SYRTE and the 100 MHz modulation after one round trip in the second fiber. Fast fluctuations are compensated by stressing a 15-meter optical fiber wrapped around a cylindrical piezo-electric actuator, which is implemented at the input of the second optical fiber. Slow perturbations are compensated by heating a 1-km optical fiber spool. In order to increase isolation between channels, the forward and return beams are modulated at different frequencies, 1 GHz and 100 MHz respectively. In this case both the one-way and the round trip signals are corrected.

These two correction systems present very similar performance and are interchangeable. The control bandwidth, about 300 Hz is mainly limited by round trip delay (0.4ms). The spectral density of the residual phase fluctuations is $-120$ dB (rad$^2$/Hz) at 1 Hz, which limits the 1s frequency stability to about $8\times10^{-15}$. The phase instability of the link without the benefits of the control system (open loop) can be simultaneously measured by slightly modulating the signal transmitted in the electronically controlled link at a frequency (270 MHz) that is different from the operating 100 MHz frequency. Thus both open and closed loop stabilities can be simultaneously recorded. The results relating to the 86km round trip link are displayed on Fig 1: the fractional frequency stability is $1.2\times10^{-14}$ at 1s, and $1\times10^{-17}$ for one day integration time. A stability of $8\times10^{-15}$ at 1 s for the one-way transfer can be deduced, since the compensation systems are independent. The long-term fluctuations, due to the daily temperature variations are reduced by two orders of magnitude. Finally the stability performance of the Cs fountain can be retrieved with no degradation at the output of the link. The 100 MHz signal short-term stability performance is slightly degraded by the optical link, but with no impact as far as frequency measurements are concerned.

As a first application of the compensated optical link, we measured the absolute frequency of a molecular optical clock. A $CO_2$ laser emitting around 30 THz is locked onto a saturated absorption line of the $OsO_4$ molecule. We chose the very strong R(42) $A_1^3(-)$ line of $^{192}OsO_4$ as it is coincident with the R(12) $CO_2$ laser line. Two equivalent and independent $CO_2/OsO_4$ stabilization systems have been developed [20]. An Allan deviation of $4.2\times10^{-14}$ at 1 s was obtained for the beat note between these two lasers (that is $3\times10^{-14}$ for a single laser). The principle of frequency measurements at 30 THz using a fs laser comb has been described elsewhere [18]. The $CO_2$ laser frequency is smaller than the span of the modes of the fs laser, and this can be exploited so that the $CO_2$ laser controls the separation between two of these modes. The stability of the $CO_2/OsO_4$ system is thus transferred to the repetition rate of the fs laser and we can dispense with any need for controlling the offset frequency of the fs laser. Two laser diodes at 852 nm and 788 nm are used as intermediate oscillators, their frequency difference being phase-locked to the $CO_2$ laser frequency. Finally $f(CO_2) = qf_r \pm \Delta$, where $\Delta$ corresponds to the offsets of the phase lock loops, and the integer q is around 29100. The $CO_2$ frequency is deduced from the measurement of the repetition rate $f_r$, which is detected with a fast photodiode. A schematic view of the experimental apparatus is given on Fig. 3. The SYRTE reference signal is used to synchronize all the synthesizers and counter at LPL, and to synthesize a 1 GHz signal, which is mixed with the repetition rate signal. The resulting beat note signal, at about 5 MHz, is further shifted to 68 Hz, filtered (10 Hz bandwidth) and counted.

Fig 4 displays the fractional Allan deviation associated with the repetition rate measurement. It is $3\times10^{-14}$ for 1 s, which demonstrates that the Allan deviation is limited by the stability of the $CO_2/OsO_4$ system. If we consider the Allan deviations at 1 s of, respectively, $8\times10^{-15}$ for the optical link, $8\times10^{-15}$ for the 100 MHz SYRTE reference signal (from the CSO), and $3\times10^{-14}$ for the $CO_2/OsO_4$ stabilized laser, we can derive a conservative upper limit of $1\times10^{-14}$ for the Allan deviation at 1 s integration time, for the equipment comparing the $CO_2/OsO_4$ signal with the microwave signal including fs laser, phase-lock loops and detection electronics. This level of uncertainty reflects a significant advance in the direct measurement of the instabilities of a fs laser-based synthesizer connecting microwave and optical domains. It compares very well with the noise level of $2\times10^{-14}$ at 1 s measured with a microwave beat note of two fs lasers driven by the same optical frequency standard [14].

The optical link to the primary standard of SYRTE allows an absolute frequency measurement, without any need for long integration time. It is remarkable indeed that the resolution of the $CO_2/OsO_4$ frequency measurement is $3\times10^{-14}$ after only 1 s integration time. The absolute frequency measurement of the R(42) line of $OsO_4$ was performed by repeating the $CO_2/OsO_4$ frequency measurement a few times. The mean value is : $\nu - \nu_{ref} = 3.9 \pm 10\ Hz$, where the uncertainty is the 1-σ deviation of the data and $\nu_{ref}$ is deduced from [21, 22]. The uncertainty is limited by the reproducibility of the $CO_2/OsO_4$ stabilization [18, 20].

We have demonstrated a transfer of the stability of a primary standard along a 43 km optical link of telecommunication network with a resolution of $8\times10^{-15}$ at 1 s and $10^{-17}$ for one day. As a demonstration, the transferred signal of a cryogenic oscillator was used to measure a $CO_2/OsO_4$ optical frequency standard, with a resolution of $3\times10^{-14}$ at 1 s limited by the $CO_2$ laser fluctuations. The long-term frequency fluctuations of this optical link are 2 orders of magnitude lower than that obtainable with a GPS-based transfer system, and the short-term fluctuations are one order of magnitude lower than the stability of a H-maser. It appears that this type of approach could provide a better frequency reference than the use of conventional local time standards.

Our next development will increase the reference frequency from 100 MHz to 1 GHz. An improvement by a factor 3 should then be obtained for the short-term resolution of the optical link. For transmission between more distant laboratories, both the fiber attenuation and the higher noise can degrade the signal transfer. Optical amplification and the use of more powerful laser diodes can overcome this attenuation, extending the performance of our compensated link for distances up to 200 km. For more distant laboratories, a scaled up version of this link using a few repeaters could further extend the range of an optical distribution system. In fact, in the present arrangement the stabilization system at LPL is acting as a repeater at the midway point of the round trip.

A very attractive application of the present work concerns the search for possible time variations of the fundamental constants [5]. Recent measurements [1, 2, 23, 24] have shown no time dependence of the fine structure constant at the limit of around $10^{-15}$/year, which is approaching the sensitivity needed to give quantitative results for the theoretical models. For the next generation of optical clocks, an even higher resolution will be needed for the frequency comparison. The present performance of our frequency dissemination system, or its extension to a continental basis, could be fully exploited in such a program. It would also allow comparison between a larger set of clocks, and thus enable discrimination between the variations of different constants [2].

ACKNOWLEDGEMENTS

The authors would like to thank the Ministère de la Recherche, the CNRS, the Bureau National de Métrologie and the European Space Agency for specific funds.

**FIGURES CAPTION**

FIGURE 1: Allan deviation for the round-trip optical link, 86 km, in open loop (▲) and closed loop (▼), as calculated from phase measurement with 1 s sample time. Dashed line indicates the fractional frequency instability of the 100 MHz reference signal from the CSO, and dotted line the fractional frequency instability of the Cs fountain FO2.

FIGURE 2: Schematic of the optical link compensation.

FIGURE 3: Schematic of the whole measurement set-up.

FIGURE 4: Fractional $CO_2/OsO_4$ frequency stability as given by the Allan deviation (●) of the repetition rate $f_r$ of the fs comb calculated from a series of 1-s gate measurement. The noise due to the optical link is negligible (▼)

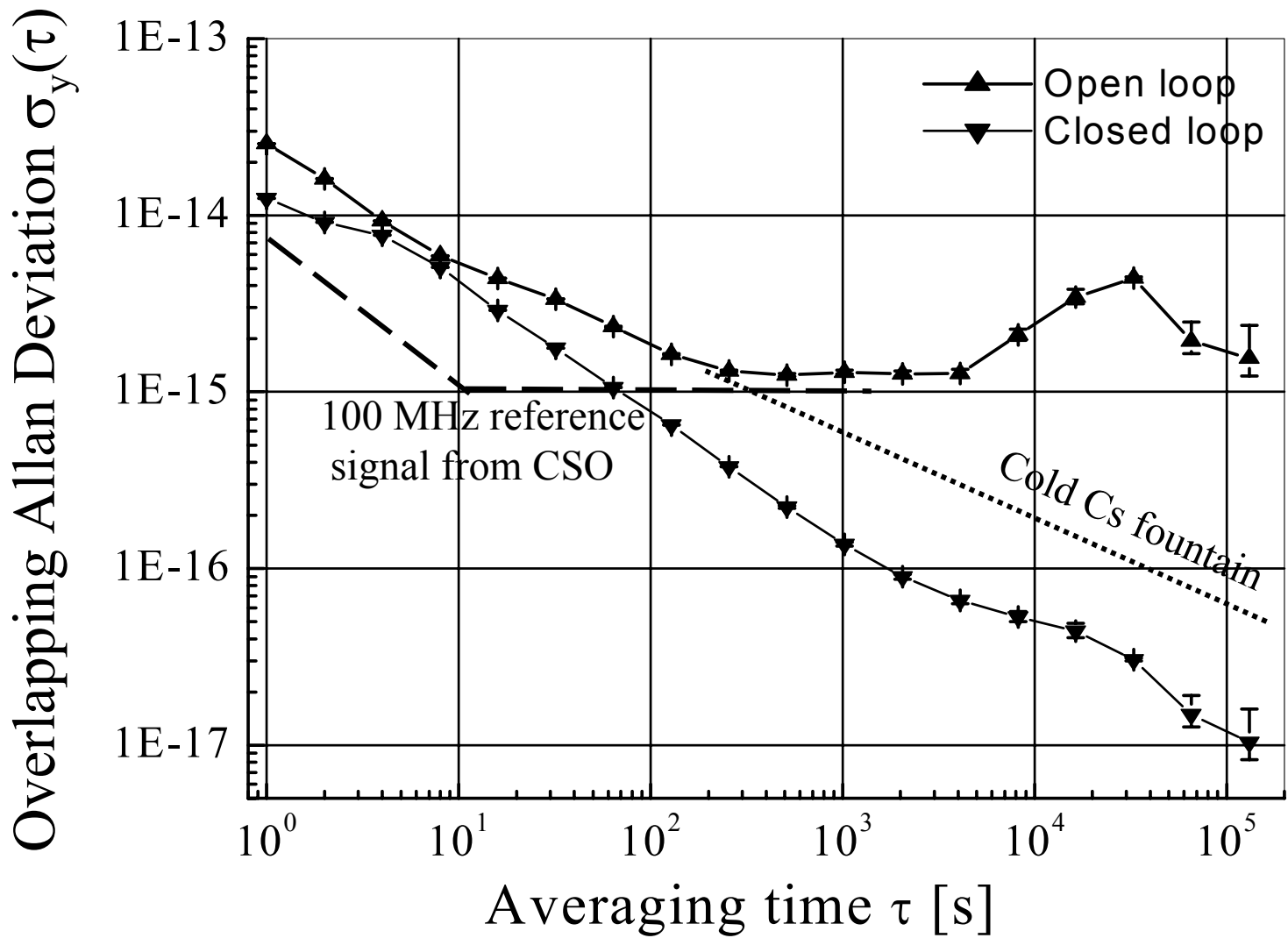

## SYRTE | LPL

**Double 43-km Optical link**

- CSO/Maser/Cs
- Ref : 100 MHz
- 100 MHz + $\Delta\phi_c$
- DL 1.55 µm
- electronic correction $\Delta\phi_c = -\Delta\psi$
- $2\Delta\psi + \Delta\phi_c$ @100 MHz
- Out
- 100 MHz $\Delta\psi + \Delta\phi_c$
- DL 1.55 µm
- 100 MHz
- × 10
- 1 GHz
- DL 1.55 µm
- 100 MHz
- Back
- DL 1.55 µm
- 1 GHz
- ÷10
- 100 MHz
- Thermal and mechanical corrections
- $\Delta\phi$
- Allan Deviation

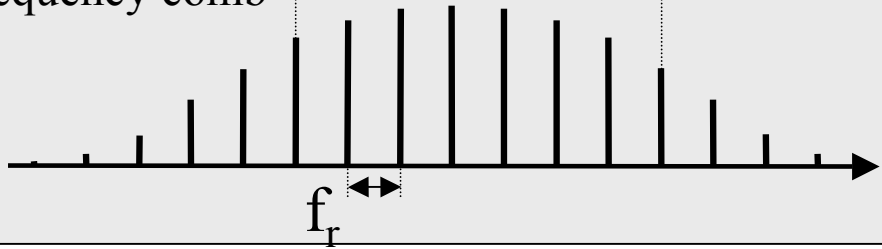

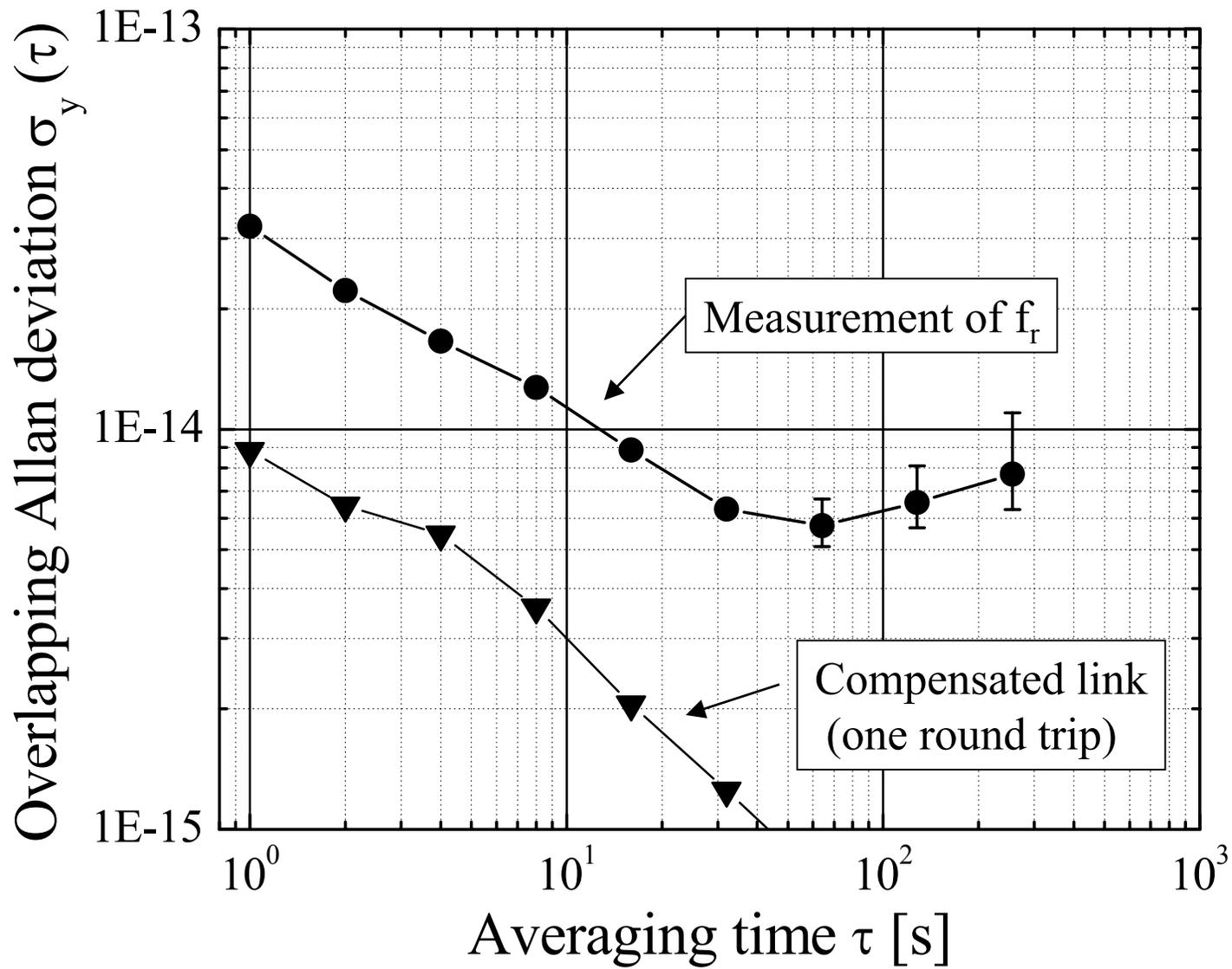